\documentclass{ws-ijmpb}

\begin{document}

\markboth{Yong Xu and Chuanwei Zhang}
{Topological Fulde-Ferrell Superfluids of a Spin-Orbit Coupled Fermi Gas}

\catchline{}{}{}{}{}
%
%%%%%%%%%%%%%%%%%%%%%%%%%%%%%%%%%%%%%%%%%%%%%%%%%%%%%%%%%%%%%%%%%%%%

\title{Topological Fulde-Ferrell Superfluids of a Spin-Orbit Coupled Fermi Gas}

\author{Yong Xu}

\address{Department of Physics, The University of Texas at Dallas, Richardson,
Texas 75080, USA \\
yongxuphy@gmail.com }

\author{Chuanwei Zhang}

\address{Department of Physics, The University of Texas at Dallas, Richardson,
Texas 75080, USA \\
chuanwei.zhang@utdallas.edu}

\maketitle

\begin{history}
\received{Day Month Year}
\revised{Day Month Year}
%\accepted{(Day Month Year)}
%\comby{(xxxxxxxxxx)}
\end{history}

\begin{abstract}
Topological Fermi superfluids have played the central role in various
fields of physics. However, all previous studies focus on the cases where
Cooper pairs have zero center-of-mass momenta (i.e. normal superfluids).
The topology of Fulde-Ferrell superfluids with nonzero momentum pairings
have never been explored until recent findings that Fulde-Ferrell superfluids
in a spin-orbit coupled Fermi gas can accommodate Majorana fermions in
real space in low dimensions and Weyl fermions in momentum space
in three dimension. In this review, we first discuss the mechanism
of pairings in spin-orbit coupled Fermi gases in optical lattices
subject to Zeeman fields, showing that spin-orbit coupling as well
as Zeeman fields enhance Fulde-Ferrell states while
suppress Larkin-Ovchinnikov states. We then present the low temperature
phase diagram including both FF superfluids and topological FF superfluids
phases in both two dimension and three dimension. In one dimension,
Majorana fermions as well as phase dependent order parameter are
visualized. In three dimension, we show the properties of Weyl fermions
in momentum space such as anisotropic linear dispersion, Fermi arch, and
gaplessness away from $k_{\perp}=0$. Finally, we discuss some possible methods to
probe FF superfluids and topological FF superfluids in cold atom systems.
\end{abstract}


\section{Introduction}
The exotic Fulde-Ferrell-Larkin-Ovchinnikov (FFLO) state
was proposed as the ground state for a superconductor with strong Zeeman splitting
about half a century ago~\cite{FuldePR1964,Larkin1964}. The Cooper pairs of such
states have finite
center-of-mass momenta with the spatially dependent order parameter, instead of zero
momenta with the spatially uniform order parameter in normal superconductors.
To date, intense search has been undertaken in solid materials such as
heavy-fermion superconductors
~\cite{HeaFermFFLO1,HeaFermFFLO2,HeaFermFFLO3,HeaFermFFLO4,%
HeaFermFFLO5},
organic superconductors~\cite{OrgSupFFLO1,OrgSupFFLO2,OrgSupFFLO3,OrgSupFFLO4},
iron pnictide superconductors~\cite{Cho2011PRB,Ptok2013JLTP}, and two-dimensional electron
gases~\cite{Li2011Nature}. Despite some remarkable findings,
conclusive evidence has not been observed. The realization of ultralcold
atom systems provides a disorder-free and highly controllable platform to
simulate quantum phenomena. In this system, FFLO states have been predicted
to exist in a polarized Fermi gas~\cite{Sheehy2006PRL,Simons2007Nature,Devreese2013Arxiv} as well as a
polarized fermionic optical
lattice~\cite{Torma2007PRL,Meisner2007PRB,qinghong2008PRB,Yan2009PRB,Trivedi2010PRL,Cai2011PRA,Wolak2012PRA},
where the polarization can be readily tuned by the particle number difference of two component fermions.
The phase diagram where the FFLO states
occupy is much larger in 1D or quasi-1D~\cite{Orso2007PRL,Hu2007PRL,Parish2007PRL,%
Vincent2008PRA,Mueller2008PRA,HuangQiang2008PRL,Ueda2008PRL,Sun11,Kuei2012PRA,MeisnerReview,Xiwen2013RMP}
than that in 3D due to a nesting
effect~\cite{Yang2001PRB}. In experiments, the Hulet group~\cite{Liao2010Nature} has measured
the density profiles of an imbalanced two component mixture of ultralcold $^{6}Li$ atoms in a
quasi-1D geometry, showing a partially polarized core surrounded by paired
or fully polarized shells, in good agreement with theoretical
calculations~\cite{Orso2007PRL,Hu2007PRL}.
However, the superfluidity of the polarized core has
not been detected, leaving the FFLO superfluids still ambiguous.

Majorana fermions, quantum particles which are their own anti-particles,
have been the focus of many theorists and experimentalists in superconductivity/superfluidity
because of their tantalizing properties
~\cite{Green2000PRB,Mizushima2008PRL,FuLiang2008PRL,%
DasSarma2008RMP,Sau2010PRL,Roman2010PRL,Oreg2010PRL,XiaoLiang2011RMP,Alicea2012RPP,%
Kouwenhoven2012Science,Xu2012Nano,Shtrikman2012Nat,Furdyna2012Nat,Yun2013NJP} and potential applications in
fault-tolerant quantum computation~\cite{Kitaev2003AP}. Majorana fermions
generally locate at the defects such as vortices, edges, and domain walls
in real space in low dimensions as zero energy quasiparticle excitations.
Interestingly, another type of topological fermions, Weyl fermions~\cite{Weyl},
can also exist as quasiparticle excitations in superfluids, not in real space
of low dimensional systems but in momentum space of three dimensional (3D) ones.
Such superfluids include $^3$He A phase~\cite{volovik}, spin-orbit
coupled Fermi gases~\cite{Gong2011prl,Sumanta2013PRA}, spin-orbit coupled Fulde-Ferrell (FF)
superfluids~\cite{Yong2014PRL,Dong2014arXiv}, and
dipolar Fermi gases~\cite{LiuBo2014arXiv}. Weyl fermions are massless chiral
Dirac fermions in 3D momentum space with linear energy dispersion.
Such fermions are robust and can only be destroyed by merging two Weyl
fermions with opposite charges, in sharp contrast to its analog in 2D,
Dirac fermions (e.g. graphene) whose gap can be
opened by the perturbation breaking time-reversal symmetry or space inversion symmetry.

The realization of spin-orbit (SO) coupling in cold atom systems
~\cite{Lin2011Nature,Jing2012PRL,Zwierlen2012PRL,PanJian2012PRL,Qu2013PRA,Spilman2013PRL,WeiArxiv}
has provided neutral atoms an opportunity to exhibit intriguing ground states
both in Boson and Fermi gases~\cite{Ohberg2011RMP,ZhaiHui2012JMPB,Spilman2013NatRev,Zhou2013AMOP},
which are remarkably different from those without the SO coupling.
In Boson gases, the ground state can be stripe or plane wave
~\cite{ZhaiHui2010PRL,Galitski2008PRA,Wu2011CPL,Yongping2012PRL,Santos2011PRL,Hu2012PRL}
for repulsive interactions and can be bright solitons with spin parity symmetry
for attractive interactions~\cite{Yong2013PRA,Achilleos2014PRL}. The shape of such bright
solitons depends on their velocity because of the breaking of Galilean invariance
~\cite{Qizhong2013,Yong2013PRA}.
These intriguing ground states are mainly due to the appearance of a double well
structure of the underlying single particle spectrum.
In Fermi gases, the ground state of superfluids with SO coupling can become
topological in the presence of out-of-plane Zeeman fields and such topological superfluids
accommodate Majorana fermions in low dimensions~\cite{Zhang2008PRL,%
Sato2009PRL,ShiLiang2011PRL,LJiang2011PRL,XJLiu2012PRA,Melo2012PRL,%
Gong2012PRL,XiongXiv2013,Yong2014Soliton} or
Weyl fermions in 3D~\cite{Gong2011prl,Sumanta2013PRA}. Recently, such SO coupled
superfluids with in-plane Zeeman splitting have been found to support Fulde-Ferrell states,
which are dominant in the low temperature phase diagram
~\cite{Zheng2013PRA,FanWu2013PRL,Zheng2013arXiv,Liu2013PRA,Fu2013PRA,LinDongArx,Hui2013PRA,%
Iskin2013,LeiJiang2014}.
The competition between Larkin-Ovchinnikov(LO) and FF states was examined
and the pairing mechanism was discussed in optical lattices~\cite{YongPRA14}.
Later, it was found that in the presence of both in-plane and out-of-plane
Zeeman fields the FF superfluids can become topological in low dimensions
as well as 3D and support Majorana fermions~\cite{Qu2013NC,Yi2013NC,XJ2013PRA,Chun2013PRL}
and anisotropic Weyl fermion excitations~\cite{Yong2014PRL,Dong2014arXiv}
respectively.

In this review, we will present some essential aspects of topological FF superfluids.
In Sec.~\ref{sec1}, we discuss the effects of Zeeman fields on the single
particle spectrum of SO coupled Fermi gases and the mechanism of FF Cooper
pairings. In Sec.~\ref{sec2}, we describe the topological properties
of FF superfluids in 1D, 2D and 3D. We also discuss the feasibility to
observe the topological FF superfluids by speeds of sound.

\section{Single Particle Picture and Pairing Mechanism}
\label{sec1}
We first consider the effects of SO coupling and Zeeman fields on
the single particle spectrum. The single particle Hamiltonian can
be written as
\begin{equation}
H_{s}(\hat{\mathbf{p}})=\frac{\hat{
\mathbf{p}}^{2}}{2m}-\mu +H_{\text{SOC}}(\hat{\mathbf{p}})+H_{z}
\end{equation}
with
momentum operator $\hat{\mathbf{p}}=-i\hbar (\partial _{x}\mathbf{e}%
_{x}+\partial _{y}\mathbf{e}_{y})$, chemical potential $\mu $, and
the atom mass $m$. The Rashba SO coupling $H_{%
\text{SOC}}(\hat{\mathbf{p}})=\alpha (\hat{\mathbf{p}}\times \mathbf{\sigma }%
)\cdot {\mathbf{e}_{z}}$ with Pauli matrix $\mathbf{\sigma }$; the Zeeman
field $H_{z}=h_{x}\sigma _{x}+h_{z}\sigma _{z}$ along the $x$ (in-plane) and
$z$ (out-of-plane) directions. In the presence of SO coupling, spin is
no longer a good quantum number, but replaced by the helical index
(eigenvalue of $((\hat{\mathbf{p}}\times \mathbf{\sigma }%
)\cdot {\mathbf{e}_{z}})/|p|$). Here we consider Rashba SO coupling, so that
the kinetic energy term along the z direction can be incorporated
into the chemical potential and this doesn't change the single particle physics
and relevant pairing mechanism. Thus here we focus on 2D.
This Hamiltonian is readily diagonalized in momentum space. It is evident that
the energy spectrum has the rotational symmetry, which cannot be broken by out-of-plane
Zeeman fields. However, in-plane Zeeman fields are capable of breaking this symmetry.
We define two operators $U_x=\sigma_y\mathcal{T}$ and $U_y=\sigma_x\mathcal{T}$ where
$\mathcal{T}$ is the time reversal operator. It can be easily seen that with an in-plane
Zeeman field ($h_x$) along the x direction, the symmetry defined by $U_x$ is satisfied
that $U_x H_s(\hat{\bf p}) U_x^{-1}=H_s(\hat{p}_x,-\hat{p}_y)$ leading to
$E(-k_x,k_y)=E(k_x,k_y)$. However, the symmetry defined by $U_y$ is broken,
generally leading to $E(k_x,-k_y)\neq E(k_x,k_y)$. When the in-plane Zeeman
field is along the y direction, the opposite is true.

\begin{figure}[bt]
\centerline{\psfig{file=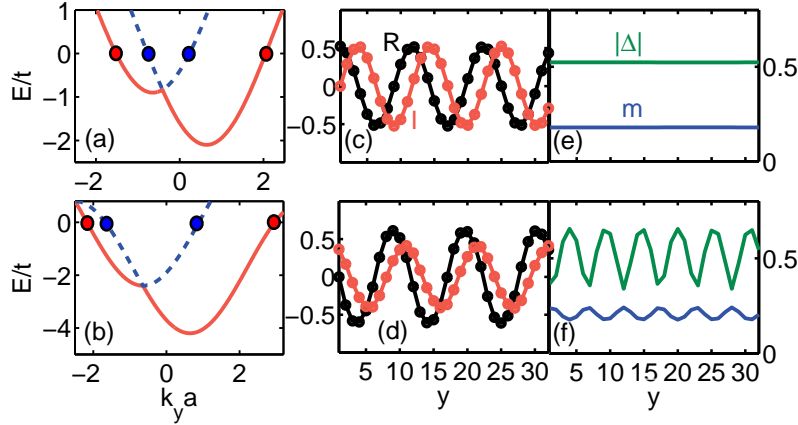,width=4.2in}}
\vspace*{8pt}
\caption{The single particle spectrum along $k_y$ with $k_x=0$
in (a) and (b), where the solid red and dashed blue lines represent the
helicity - and helicity + branches respectively. The real and imaginary parts of
the order parameter in (c) and (d). The absolute value of the order parameter and the magnetism
in (e) and (f). The first row corresponds to a FF state, while the second
a LO state.}
\label{single_particle}
\end{figure}

To be specific, the energy spectrum reads
\begin{equation}
E({\bf k})=\frac{\hbar^2 {\bf k}^2}{2m}\pm\sqrt{\alpha^2 k_x^2+(h_x-\alpha k_y)^2+h_z^2}.
\end{equation}
which clearly exhibits the symmetry breaking in the presence of $h_x$. The effect of
$h_z$ is to open a gap between two helical branches whereas $h_x$ cannot as shown in
Fig.~\ref{single_particle}(a,b).

In the following, we discuss the pairing issue in the presence of
attractive interactions. One question is whether the pairing mainly occurs
around the Fermi surface in the BCS regime, leading to finite center-of-mass
momentum Cooper pairings (i.e. FF superfluids) because of the distortion of the single
particle spectrum. Another question is whether FF superfluids
are really the ground states instead of LO superfluids.
FF and LO states are two types of FFLO states associated with phase dependent order parameter
$\Delta({\bf r})\propto e^{i{\bf Q}\cdot {\bf r}}$
and with amplitude oscillating
order parameter $\Delta({\bf r})\propto\text{cos}({\bf Q}\cdot{\bf r})$, respectively.
LO states can be regarded as the interference states of two Cooper
pairs (two FF states) with opposite center-of-mass momenta. In a system with the
inversion symmetric Fermi surface, LO states are generally energetically
favorable than FF states because if there are Cooper pairs with momenta
${\bf Q}$, the existence of Cooper pairs with momenta $-{\bf Q}$ can
decrease the energy~\cite{Trivedi2010PRL}.

To address these crucial questions,
we perform the calculation of Bogoliubov-de Gennes (BdG) equations
self-consistently in the
optical lattice model to gain the
order parameter and the pairing density in the helical representation,
since in the real space model
it is not necessary to postulate the form of Cooper pairs.
In optical lattices, based on
tight-binding model, the Hamiltonian in momentum space can be obtained by
replacing the kinetic energy and $H_{SOC}$ with
$-2t(\text{cos}(k_x a)+\text{cos}(k_y a))$,
$\alpha (\text{sin}(k_x a)\sigma_y-\text{sin}(k_y a)\sigma_x)$, respectively.
Here $t$ and $a$ are the hopping parameter and lattice constant, depending
on the strength and wavelength of laser beams respectively.

\begin{figure}[bt]
\centerline{\psfig{file=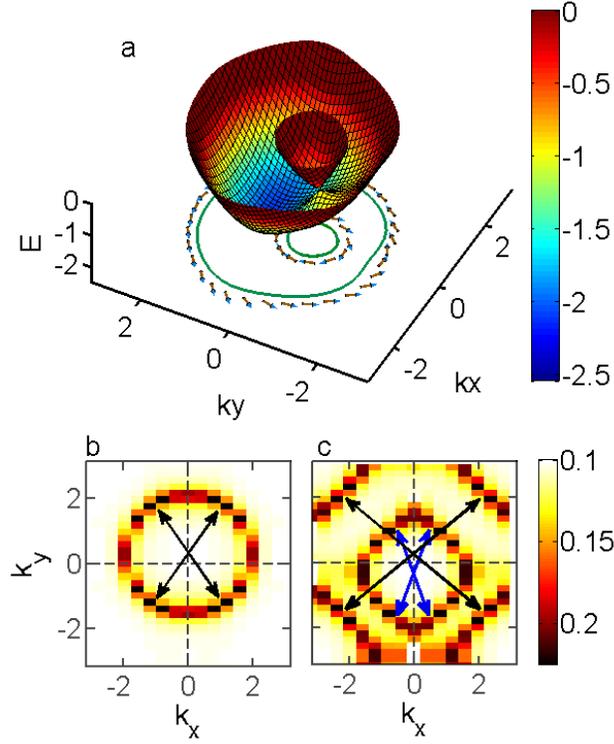,width=3.2in}}
\vspace*{8pt}
\caption{(a) Single particle band structure in momentum
space for a specific FF state. The
Fermi surface is plotted on the bottom layer with green line. The small
arrows around the Fermi surface are the spin orientations. (b) shows the
pairing density $|\langle \hat{c}_{\mathrm{\mathbf{k}},-}\hat{c}_{-\mathrm{%
\mathbf{k}}+Q_{y},-}\rangle |^{2}$ for a FF state, whose single particle spectrum
is plotted in (a); here $-$ indicates the lower branch. The black double
arrows show the pairing. (c) presents the
pairing density $|\langle \hat{c}_{\mathrm{\mathbf{k}},-}\hat{c}_{-\mathrm{%
\mathbf{k}}+Q_{y},-}\rangle |^{2}$, and $|\langle \hat{c}_{\mathrm{\mathbf{k}%
},+}\hat{c}_{-\mathrm{\mathbf{k}}-Q_{y},+}\rangle |^{2}$ for the LO phase.
Here $+$ indicates the higher branch. The black and blue double arrows illustrate the
Cooper pairings with $Q_{y}$ and $-Q_{y}$, respectively. $Q_{y}$ depends on
the deformation of the Fermi surface by the SO coupling and the Zeeman
field. The units of $E$, and $k_{x}$, $k_{y}$ are $t$ and $1/a$ respectively.
From Ref.~\protect\refcite{YongPRA14}.}
\label{FF_pairing}
\end{figure}

To examine whether fermionic atoms form Cooper pairs around the Fermi surface
in the same helical branch, Fig.~\ref{FF_pairing}(b,c) plots the pairing
density $|\langle \hat{c}_{{\bf k},-}\hat{c}_{-\mathrm{%
\mathbf{k}}+Q_{y},-}\rangle |^{2}$ and
$|\langle \hat{c}_{\mathrm{\mathbf{k}%
},+}\hat{c}_{-\mathrm{\mathbf{k}}-Q_{y},+}\rangle |^{2}$,
where $\hat{c}_{{\mathrm{\mathbf{k}}},\lambda}$
annihilates an atom with momentum ${\bf k}$ in the helical $\lambda$ branch.
It shows that Cooper pairs are mainly formed by the atoms around the Fermi
surface in the same helical branch. The pairs possess finite momenta
(thus FFLO superfluids) due to
the distortion of band structure along the y direction as shown in
Fig.~\ref{FF_pairing}(a).
In FF states, only the helical - branch
takes part in the pairing while the contribution of the helical + branch is
extremely small because of much smaller density of states. In contrast,
atoms in both branches participate in the pairing for LO states.

Based on the above results, FF states are not always the ground state and
LO states still have the chance to be. It mainly depends on the single particle
structure and the chemical
potential (See phase diagram in Ref.~\refcite{YongPRA14}).
In Fig.~\ref{single_particle}, we take two examples associated with FF
and LO states respectively to illustrate these effects.
Fig.~\ref{single_particle}(a,b) presents the single particle energy with
respect to $k_y$ for fixed $k_x=0$.
%It is clearly shown that $h_x$ distorts
%the energy spectrum along the y direction but does not open the gap
%between two helical branches.
In the right panel, the superfluids order
parameter $\Delta(i)$ calculated self-consistently in Ref.~\refcite{YongPRA14}
is plotted. Here
$\Delta(i)=U\langle \hat{c}_{i\downarrow}\hat{c}_{i\uparrow}\rangle$
with interaction strength $U$, annihilation operator $\hat{c}_{i\sigma}$
at $i$ site with spin $\sigma$. (c) and (e) have FF type pairing with
space dependent phase order parameter while (d) and (f) LO type with
space oscillating amplitude order parameter.
Since the pairing happens in the same helical branch as red and blue
points in Fig.~\ref{single_particle}(a,b), a local superfluid order parameter can be
written as $\Delta _{i}=\Delta _{-}\exp \left( iyQ_{y-}\right) +\Delta
_{+}\exp \left( iyQ_{y+}\right) $ with the former part contributed by
the paring in the helical - branch and the latter one in the helical
+ branch. When the chemical potential is placed where the density of
states in the helical + branch is small, $\Delta_{+}\approx 0$ and
$\Delta _{i}=\Delta _{-}\exp \left( iyQ_{y-}\right)$, the FF
pairing, as shown in the first row of Fig.~\ref{single_particle}; however,
when the opposite is true as shown in the second row of
Fig.~\ref{single_particle}, $\Delta_{+}\neq 0$ and $Q_{y-}\approx -Q_{y+}$,
leading to a LO superfluid. It is important to note that such LO states
are different from the traditional LO states where the order parameter
is real and has nodes. Because $|\Delta_{-}|>|\Delta_{+}|$ due to higher
density of states in the helical - branch, such generalized LO states are
complex and have the nonzero order parameter domain walls. In general,
the SO coupling and in-plane Zeeman fields enhance the FF superfluids and
suppress the LO superfluids.

\section{Topological FF superfluids in SO Coupled Fermi Gases }
\label{sec2}
\subsection{Model and Effective Hamiltonian}
We consider a SO coupled Fermi gas subject to both in-plane and out-of-plane
Zeeman fields and \textit{s}-wave contact interactions. The many-body
Hamiltonian can be written as
\begin{equation}
H=\int d\mathbf{r}\hat{\Psi}^{\dagger }(\mathbf{r})H_{s}(\hat{%
\mathbf{p}})\hat{\Psi}(\mathbf{r})-U\int d\mathbf{r}\hat{\Psi}_{\uparrow
}^{\dagger }(\mathbf{r})\hat{\Psi}_{\downarrow }^{\dagger }(\mathbf{r})\hat{%
\Psi}_{\downarrow }(\mathbf{r})\hat{\Psi}_{\uparrow }(\mathbf{r}),
\end{equation}
where $U$ characterizes the strength of attractive interactions;
$\hat{\Psi}(\mathbf{r}%
)=[\hat{\Psi}_{\uparrow }(\mathbf{r}),\hat{\Psi}_{\downarrow }(\mathbf{r}%
)]^{T}$ and $\hat{\Psi}_{\nu }^{\dagger }(\mathbf{r})$ ($\hat{\Psi}_{\nu }(%
\mathbf{r})$) is fermionic atom creation (annihilation) operator.

In quantum field theory, the partition function can be written as $Z=\text{Tr%
}(e^{-\beta H})=\int D(\bar{\psi},\psi )e^{-S_{eff}[\bar{\psi},\psi ]}$ with
$\beta =1/k_{B}T$. The effective action is
\begin{equation}
S_{eff}[\bar{\psi},\psi ]=\int_{0}^{\beta }d\tau \left( \int d\mathbf{r}%
\sum_{\sigma }\bar{\psi}_{\sigma }(\mathbf{r},\tau )\partial _{\tau }\psi
_{\sigma }(\mathbf{r},\tau )+H(\bar{\psi},\psi )\right) ,
\end{equation}%
where $\int d\tau $ is an integral over the imaginary time $\tau $ and $H(%
\bar{\psi},\psi )$ is obtained by replacing $\hat{\Psi}_{\sigma }^{\dagger }$
and $\hat{\Psi}_{\sigma }$ with Grassman field number $\bar{\psi}_{\sigma }$
and $\psi _{\sigma }$. The quartic interaction term is transformed to
quadratic one by Hubbard-Stratonovich transformation, where the order
parameter $\Delta (\mathbf{r},\tau )$ is defined. By integrating out fermion
fields, the partition function reads $Z=\int D(\bar{\Delta},\Delta
)e^{-S_{eff}[\bar{\Delta},\Delta ]}$, where the effective action can be
written as
\begin{equation}
S_{eff}[\bar{\Delta},\Delta ]=\int_{0}^{\beta }d\tau \int d\mathbf{r}(\frac{%
|\Delta |^{2}}{U})-\frac{1}{2}\ln \det {G}^{-1}.
\end{equation}%
Here the inverse single particle Green function $G^{-1}=-\partial _{\tau
}-H_{B}({\hat{\bf p}})$ in the Nambu-Gor'kov representation with $4\times 4$ Bogoliubov-de
Gennes (BdG) Hamiltonian
\begin{equation}
H_{B}({\hat{\bf p}})=\left(
\begin{array}{cc}
H_{s}(\hat{\mathbf{p}}) & \Delta (\mathbf{r},\tau ) \\
\Delta (\mathbf{r},\tau ) & -\sigma _{y}H_{s}(\hat{\mathbf{p}})^{\ast }\sigma _{y}
\end{array}%
\right) .
\end{equation}%
Based on the conclusion in the previous section that SO coupling and in-plane
Zeeman fields enhance the FF states, we assume the FF form order parameter of
Fermi gases, $\Delta (\mathbf{r},\tau
)_{0}=e^{iQ_{y}y}\Delta _{0}$ with the space independent $\Delta _{0}$.
By Fourier transformation and
summing the Matsubara frequency, this form of $\Delta (\mathbf{r},\tau )$ yields
the thermodynamical potential
\begin{eqnarray}
\Omega &=&|\Delta |^{2}/U+\sum\nolimits_{\mathbf{k}}\left( \hbar ^{2}(-%
\mathbf{k}+\mathbf{Q}/2)^{2}/2m-\mu \right) \\
&&-\sum\nolimits_{\mathbf{k},\sigma }\frac{1}{2\beta }\ln (1+e^{-\beta E_{%
\mathbf{k}\sigma }}).  \notag
\end{eqnarray}%
Here $E_{\mathbf{k}\sigma }$ is the eigenvalue of $4\times 4$ Bogoliubov-de
Gennes (BdG) Hamiltonian
\begin{equation}
H_{B}=\left(
\begin{array}{cc}
H_{s}(\mathbf{k}+\mathbf{Q}/2) & \Delta _{0} \\
\Delta _{0} & -\sigma _{y}H_{s}(-\mathbf{k}+\mathbf{Q}/2)^{\ast }\sigma _{y}%
\end{array}%
\right) ,
\end{equation}%
$\mathbf{Q}=Q_{y}\mathbf{e}_{y}$ is the total momentum of the Cooper pair.
The mean-field solutions of $\Delta _{0}$, $Q_{y}$, and $\mu $ satisfy the
saddle point equations $\partial \Omega /\partial \Delta _{0}=0$, $\partial
\Omega /\partial Q_{y}=0$, and the atom number equation $\partial \Omega
/\partial \mu =-n$ with a fixed total atom density $n$. To regularize the
ultra-violet divergence at large $\mathbf{k}$, we follow the standard
procedure~\cite{StringariRMP}: in 3D $\frac{1}{U}=\frac{m}{4\pi \hbar ^{2}a_{s}}%
-\int \frac{d\mathbf{k}}{(2\pi )^{3}}\frac{m}{\hbar ^{2}k^{2}}$ with the
s-wave scattering length $a_{s}$ and in 2D $1/U=\sum_{\mathbf{k}}1/(\hbar
^{2}k^{2}/m+E_{b})$ with the binding energy $E_{b}$.
The self-consistent solution is obtained
through the minimization of the free energy $F=\Omega +\mu n$. The energy
unit is chosen as the Fermi energy $E_{F}=\hbar ^{2}\mathbf{K}_{F}^{2}/2m$
of non-interacting Fermi gases without SO coupling and Zeeman fields with
Fermi vector $K_{F}=(3\pi ^{2}n)^{1/3}$ in 3D and $K_{F}=(2\pi n)^{1/2}$ in 2D.

Such BdG Hamiltonian possesses the particle-hole symmetry $\Xi=\Lambda\mathcal{K}$
with $\Lambda=i\sigma_y\tau_y$, and the complex conjugate operator $\mathcal{K}$.
Here $\Xi^2=1$. This leads to $E_{-\mathbf{k}\sigma }=-E_{\mathbf{k}\bar{\sigma} }$
where $\bar{\sigma}=5-\sigma$ if we label the eigenvalues from 1 to 4.

\subsection{FF Superfluids}
\begin{figure}[bt]
\centerline{\psfig{file=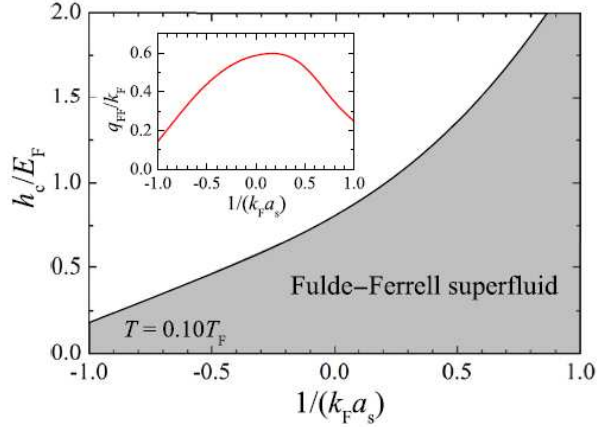,width=3.2in}}
\vspace*{8pt}
\caption{Phase diagram with respect to in-plane Zeeman fields $h_c$ across BCS-BEC
crossover at low temperature $T=0.1T_F$ (from Ref.~\protect\refcite{Hui2013PRA}).
}
\label{phase_NoT}
\end{figure}

The mean-field ground state of a SO coupled Fermi gas subject to pure
in-plane Zeeman fields was first calculated in 3D in Ref.~\refcite{Zheng2013PRA}
and then in 2D in Ref.~\refcite{FanWu2013PRL}. It was found that $Q_{y}$ is nonzero
with nonzero Zeeman fields, implying the existence of FF parings.
$Q_{y}$ is also a monotonously
increasing function of Zeeman fields. The phase diagram where FF states exist
becomes dominant as shown in Fig.~\ref{phase_NoT}, in sharp contrast to
traditional FFLO superfluids which can
only exist in a tiny fraction of the phase diagram. The free energy difference
between FF states and excited BCS states can be as large as $0.04E_F$~\cite{Zheng2013PRA}.
The gapless FF superfluids with gapless quasiparticle excitations,
originating from the strong distortion of the energy excitations,
have also been found~\cite{FanWu2013PRL,LinDongArx}. The finite temperature mean-field
phase diagram is obtained in Ref.~\refcite{LinDongArx,Hui2013PRA},
showing that FF states also dominate at finite temperature.

\subsection{Topological FF Superfluids in 1D and 2D}
In the presence of out-of-plane Zeeman fields ($h_z$), the gap between two
helical branches is opened and superfluids can become topological, which
host Majorana fermions in 1D and 2D. In the following, we first briefly
introduce the current progress of Majorana fermions and then discuss them
in FF superfluids.
\subsubsection{Majorana fermions}
Majorana fermions, quantum particles which are their own anti-particles,
have attracted tremendous interests since the Kitaev's seminal
paper~\cite{Kitaev2001} suggested their potential applications in fault-tolerant
quantum computation in 1D spinless $p$-wave superconductors. To date,
there are lots of proposals to create Majorana fermions including
2D $p_x$+$ip_y$ superconductors~\cite{Green2000PRB,Mizushima2008PRL},
the surface of 3D topological insulators in conjunction with the
proximity effect of superconductors~\cite{FuLiang2008PRL}, 1D nana wire
with SO coupling~\cite{Sau2010PRL,Roman2010PRL,Oreg2010PRL}, and
time-reversal invariant $d$ wave
superconductors~\cite{Law2012PRB,Fan2013PRL,Nagaosa2013PRL,Berg2013PRL,XJLiu2014PRX}.

In ultralcold atom systems, there are also many proposals to generate Majorana
fermions ranging from 1D SO coupled Fermi gases where Majorana fermions can
exit at different phase boundary~\cite{XJLiu2012PRA} or inside a soliton~\cite{Yong2014Soliton}
to SO coupled, staggered, and shaken optical lattices~\cite{Qu2014PRA,BoLiuMF,Zheng2014arXiv}.
Here the superfluidity is generated by the intrinsic attractive interactions, not by the proximity
effects. Recent experiments have shown the existence of zero energy peaks (i.e.
Majorana fermions)~\cite{Kouwenhoven2012Science,Xu2012Nano,Shtrikman2012Nat,Furdyna2012Nat},
however, whether the zero energy excitation
is caused by Majorana fermions or disorder~\cite{LiuJie2012PRL} is still under debate. In this
aspect, cold atom systems are an ideal platform because of their
disorder-free properties.

All the superconductors/superfluids belong to BCS pairing with zero
center-of-mass momentum Cooper pairings. Ref.~\refcite{Qu2013NC}--\refcite{Chun2013PRL}
suggest that FF superfluids can also become topological, supporting Majorana fermion
excitations in 1D and 2D.

\subsubsection{Signature of Majorana fermions in momentum space in 2D}
We first consider a 2D Rashba SO coupled Fermi gas subject to a pure out-of-plane
Zeeman field. By diagonalizing the BdG Hamiltonian in momentum space, the
quasiparticle excitations can be written as
\begin{equation}
E_{\pm}^{\lambda}({\bf k})=\lambda\sqrt{\xi_{\bf k}^2+\alpha^2k^2+h_z^2+|\Delta_0|^2 \pm
2\sqrt{h_z^2(\xi_{\bf k}^2+|\Delta_0|^2)+\alpha^2 k^2\xi_{\bf k}^2}},
\end{equation}
where $\lambda=\pm$ indicating the particle and hole branches,
$\xi_{\bf k}=\hbar^2{\bf k}^2/2m-\mu$, and $k=\sqrt{k_x^2+k_y^2}$.
Without SO coupling and Zeeman fields $(\alpha=h_z=0)$, it becomes the typical BCS
superfluids excitations $E_{{\bf k},\pm}^\lambda=\lambda\sqrt{\xi_{\bf k}^2+|\Delta_0|^2}$
with the gap $|\Delta_0|$. To obtain the gap close point in the presence of SO
coupling and Zeeman fields, which generally indicates a topological phase transition, we write the
multiplication of two particle branches of the quasiparticle excitations:
\begin{equation}
E_{+}^{+}({\bf k})E_{-}^{+}({\bf k})
=\left(h_z^2+\alpha^2k^2-\xi_{\bf k}^2-|\Delta_0|^2\right)^2+4\alpha^2k^2|\Delta_0|^2.
\end{equation}
Clearly, the gap can only close at $k=0$ and $h_z=\sqrt{\mu^2+|\Delta_0|^2}$.
Although the gap closing is not a sufficient condition for the topological transition,
here it really indicates the transition with a sharp change of Chern number.
This point also corresponds to a sharp change
of the topological index ${\mathcal M}=\text{sign}(\text{Pf}(\Gamma))$~\cite{Sau2010PRB},
where $\text{Pf}$ is the Pfaffian of the skew matrix $\Gamma=H_{B}(0)\Lambda$ with
$\Lambda=i\sigma_y\tau_y$. The appearance of the topology in a $s$-wave
interacted Fermi gas can be understood from the analogy to a spinless $p$
wave Hamiltonian. In the helical representation,
the effective Hamiltonian reads
\begin{equation}
H_{BdG}=\sum_{{\bf k}}E_{-}({\bf k})\hat{c}_{{\bf k}-}^{\dagger}\hat{c}_{{\bf k}-}
+\left(\Delta({\bf k})\hat{c}_{-{\bf k}-}\hat{c}_{{\bf k}-}+h.c.\right),
\end{equation}
where $\Delta({\bf k})=\alpha \Delta_0^* k e^{i\theta_{\bf k}}
/(2\sqrt{h_z^2+\alpha^2 k^2})$ with
the angle $\theta_{\bf k}$ between ${\bf k}$ and $k_y$. This Hamiltonian
is formally equivalent to a spinless $p_x+ip_y$ superconductor
~\cite{Green2000PRB,Mizushima2008PRL}.

\begin{figure}[bt]
\centerline{\psfig{file=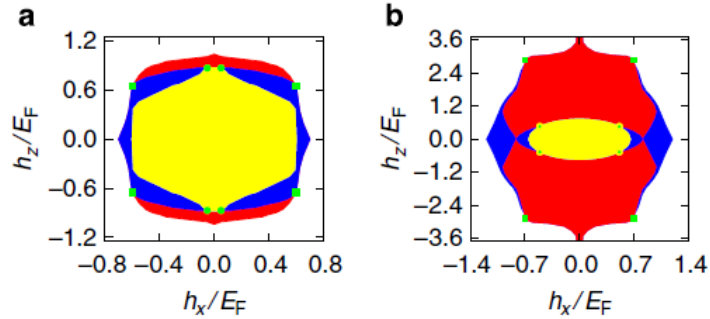,width=3.8in}}
\vspace*{8pt}
\caption{Phase diagram of FF superfluids. Red, yellow, blue, and
white regions correspond to topological gapped FF superfluids, non-topological
gapped FF superfluids, gapless FF superfluids, and normal gases, respectively.
In (a), $E_b=0.4E_F$,
$\alpha K_F=0.5E_F$; in (b), $E_b=0.4E_F$, $\alpha K_F=1.0E_F$.
From Ref.~\protect\refcite{Qu2013NC}.
}
\label{2D_phase}
\end{figure}

With in-plane Zeeman fields, Ref.~\refcite{Qu2013NC}--\refcite{XJ2013PRA}
found that topological FF superfluids can exist when
\begin{equation}
{h_z}^2+\bar{h}_x^2>\bar{\mu}^2+\Delta_0^2,\text{  }\alpha h_z\Delta_0\neq 0,\text{  }E_g>0,
\end{equation}
where $\bar{h}_x=h_x+\alpha Q_y/2$, $\bar{\mu}=\mu-Q_y^2/8m$, and $E_g$ is
the gap of the particle branches of the quasiparticle excitations. We note that this
condition was generalized to the case with $E_g<0$~\cite{Gong2014PRB,Hu2014arXiv}
after the gapless FF topological superfluids were found in 3D~\cite{Yong2014PRL}.
The nonzero value of $Q_y$ indicates the superfluids are FF type.
When $h_x=0$ thus $Q_y=0$, this condition reduces to $h_z>\sqrt{\mu^2+\Delta_0^2}$,
the forementioned requirement for the topological phase transition. With increasing
$h_x$, the critical $h_z$ decreases. However, this decreasing cannot help enhance
their Berezinskii-Kosterlitz-Thouless (BKT) temperatures~\cite{YongBKT}.
The phase diagram in Fig.~\ref{2D_phase} presents
the topological FF superfluids region (red area) in the $(h_x,h_z)$ plane, which
increases dramatically as SO coupling is enlarged.

\begin{figure}[bt]
\centerline{\psfig{file=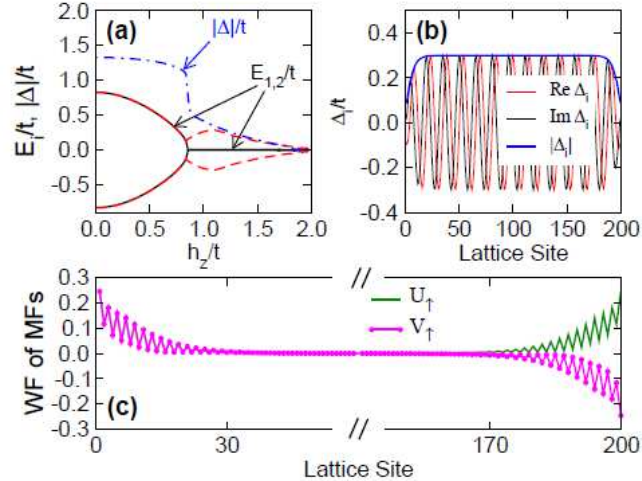,width=3.4in}}
\vspace*{8pt}
\caption{(Color online)
(a) Plot of the quasiparticle excitation energies and the
order parameter. (b) The spatial profile of the FF type order
parameter. (c) The wave function of the Majorana zero modes.
From Ref.~\protect\refcite{Qu2013NC}.}
\label{MM_MFs}
\end{figure}

\subsubsection{Realspace visualization of Majorana fermions in 1D}
Generally, Majorana fermions locate at the defects, such as vortices, edges,
and domain walls~\cite{Green2000PRB,Mizushima2008PRL,%
LJiang2011PRL,XJLiu2012PRA,Yong2014Soliton}.
To visualize the existence of zero Majorana fermion
excitations, Ref.~\refcite{Qu2013NC,Chun2013PRL} performed the
self-consistent calculation of 1D optical lattice BdG equations with the
open boundary condition. Fig.~\ref{MM_MFs} demonstrates that zero Majorana fermion
excitations, which are protected by a large gap, appear at a sufficiently
large Zeeman field and the order parameter has FF type. The wave functions
of Majorana fermions are local edge states, satisfying the self-Hermite condition.

Compared with the engineering of Majorana fermions in solid materials where
the superconductivity is generally generated through proximity effects,
the superfluid order parameter is created by intrinsic attractive interactions. The problem is
that such low dimensional systems cannot undergo conventional phase transition to
a state with long-range order,
raising the concerns whether Majorana fermions can exist in 1D and 2D Fermi gases.
In 1D, Ref.~\refcite{Roman2011PRB,Sau2011PRB} examined this problem and found
that Majorana fermions can also exist in a 1D system with algebraically decaying
superconducting fluctuations. Fluctuations may also be suppressed
in a quasi-1D system, where the long range order can be
restored by the existence of other dimensions albeit small~\cite{Parish2007PRL}.

In 2D, the relevant physics is the the BKT
transition~\cite{Berezinskii1971JETP,ThoulessJPC1972} to a state with
quasi-long-range order (i.e. vortex-antivortex (V-AV) pairs)
~\cite{Melo2006PRL,He2012PRL,Gong2012PRL,Devreese2014PRL},
with the critical temperature determined by the superfluid density tensor.
Ref.\refcite{Gong2012PRL} suggested that Majorana fermions can only be observed at finite
temperature in the sense that the distance between the vortex and antivortex
in a V-AV pair is extremely small at zero temperature leading to large
interactions between Majorana fermions locating at V-AV pairs. On the other hand,
in traditional Zeeman induced FF superfluids, the transverse superfluid density is
zero due to the rotational symmetry of the Fermi
surface~\cite{Vishwanath2009PRL,Torma2014PRB}, resulting in the zero
critical temperature. However, Ref.~\refcite{YongBKT}
found nonzero and large critical temperatures for FF superfluids,
gapless FF superfluids, topological FF superfluids, and gapless
topological FF superfluids in SO coupled Fermi gases. This paves the
way for the experimental observation of 2D gapped and gapless
FF superfluids and their associated topological excitations at
finite temperature.

\subsection{Topological FF superfluids in 3D}
\subsubsection{Weyl fermions}

Weyl fermions \cite{Weyl} are massless chiral Dirac
fermions with linear energy dispersions in momentum space, which
can be described by Weyl equations:
\begin{equation}
H({\bf k})=\pm v{\bf k}\cdot\sigma,
\end{equation}
where $\pm$ indicates the chirality of Weyl fermions. The masslessness
can be understood from the fact that no fourth matrix can be found which
anti-commutes with other three Pauli matrices. The spin expectation
distribution of the occupied hole branch exhibits a hedgehog
structure $\langle\sigma\rangle=\mp {\bf k}/k$ corresponding to the
right handed (left handed) chirality.

Weyl fermions were first proposed for describing massless
chiral Dirac fermions such as neutrinos in particle physics
in 1929. Despite much effort, these fermions have not
yet been observed in experiments (neutrinos have mass).
Recently, Weyl fermions have been suggested to exist in some
solid state materials (i.e. Weyl semimetals),
such as Pyrochlore Iridates~\cite{Wan2011prb,Aji2012prb},
ferromagnetic compound HgCr$_{2}$Se$_{4}$~\cite{ZhongFang2011prl},
multilayer topological insulators~\cite{Burkov2011PRL}, photonic crystals~%
\cite{LingLu2013NP}, as well as in optical lattices~\cite%
{Bercioux2009PRA,Lan2012PRB}. These materials possess the band
touching points, around which the energy dispersions are linear.
Such band touching points have to appear in pairs with opposite
topological charge (i.e. chirality). Compared with two dimensional
Dirac fermions (e.g., graphene), whose gap can be opened by
perturbations that break time-reversal or spatial inversion
symmetries, the gap of Weyl nodes cannot be opened unless two Weyl fermions
with opposite topological charges are merged.

In solid materials, Weyl fermions exist in a single particle spectrum.
It is natural to ask whether Weyl fermions can also exist in the
quasiparticle excitation spectrum. The pioneer work in this issue is
the proposal of Weyl fermions in $^{3}$He A phase by G. E. Volovik~\cite{volovik}.
However, $^{3}$He is a strongly interacting system and very complicated
to understand. Recently, Weyl fermions have been found in the quasiparticle
excitation spectrum of SO coupled Fermi
superfluids with simple $s$-wave interactions~\cite{Gong2011prl,%
Sumanta2013PRA}, SO coupled FF superfluids~\cite{Yong2014PRL,%
Dong2014arXiv}, nodal phases of Cu$_x$Bi$_2$Se$_3$~\cite{Yang2014PRL},
and dipolar Fermi gases~\cite{LiuBo2014arXiv}.

\subsubsection{BCS-BEC crossover and phase diagram}

\begin{figure}[bt]
\centerline{\psfig{file=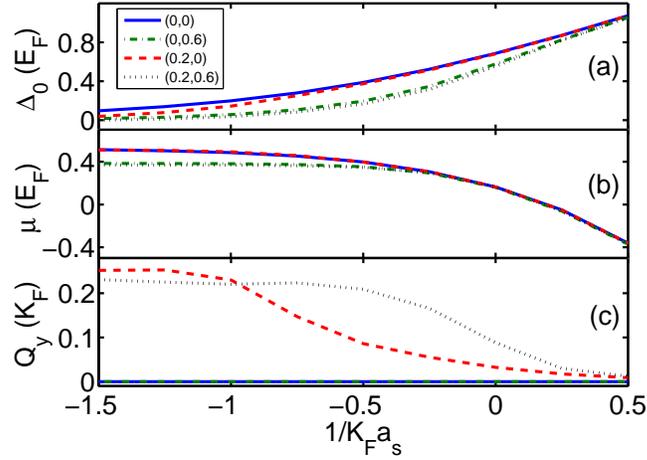,width=3.4in}}
\vspace*{8pt}
\caption{(Color online) Plot of the order parameter $\Delta _{0}$ in (a),
chemical potential $\protect\mu $ in (b), and $Q_{y}$ in (c) as a function
of $1/K_{F}a_{s}$ for different parameters (%
$h_{x}$,$h_{z}$). $\protect\alpha K_{F}=E_{F}$ and the temperature $T=0$.
From Ref.~\protect\refcite{Yong2014PRL}.}
\label{cross}
\end{figure}

Before we discuss the properties of Weyl fermions in a SO coupled Fermi gas,
we first study the BCS-BEC crossover and then map out the zero temperature
phase diagram. Fig.~\ref{cross} presents the order parameter $\Delta_0$,
chemical potential $\mu$, and $Q_y$ with respect to $1/K_F a_s$.
$\Delta_0$ increases while $\mu$ decreases with increasing
$1/K_F a_s$, leading to the same value independent of Zeeman fields
$(h_x,h_z)$ as the attractive interactions are sufficiently strong.
This signals the crossover from BCS superfluids to tightly bound molecule
BEC superfluids. The finite momentum $Q_y$ is nonzero when $h_x\neq 0$, indicating
the superfluids are FF type. This $Q_y$ is roughly a monotonously decreasing
function of $1/K_F a_s$ in the sense that the momenta of Cooper pairs induced from
weak interactions reflect the Fermi surface structure.
Both Zeeman fields are detrimental to BCS paring, leading
to the decreased $\Delta_0$ when the total Zeeman field $h=\sqrt{h_x^2+h_z^2}$ is
larger. Since the physical quantities are independent of Zeeman fields on the BEC side,
we focus on the FF superfluids on the BCS side.

\begin{figure}[bt]
\centerline{\psfig{file=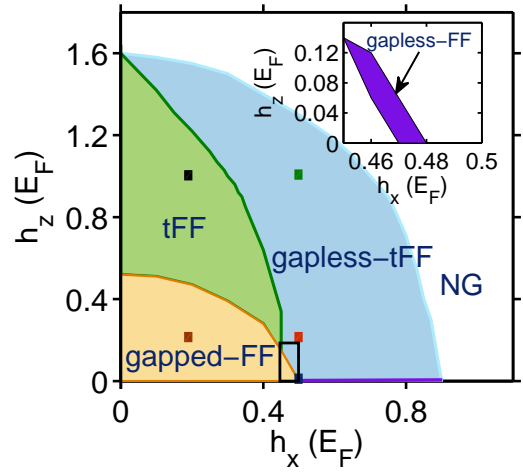,width=2.8in}}
\vspace*{8pt}
\caption{(Color online) Zero temperature mean-field phase diagram of 3D SO coupled Fermi gases
in the plane $(h_x,h_z)$. The area in the small black box is enlarged in the inset,
which shows the gapless FF phase. tFF: topological FF states; gapless-tFF: gapless topological
FF state; NG: normal gas. Here $\protect\alpha K_{F}=E_{F}$ and $1/a_{s}K_{F}=-0.1$.}
\label{phase}
\end{figure}

In Fig.~\ref{phase}, the zero temperature phase diagram is mapped out in
the plane $(h_x,h_z)$. With pure $h_x$ corresponding to the horizonal axis,
the ground state is FF state when $h_x$ is nonzero.
Such FF states can be divided into two groups: gapped one with $E_g>0$ and
gapless one with $E_g<0$, where the gap $E_g=\text{min}(E^{+}_{-})$ is defined as
the minimum of the particle branch of the quasiparticle energy excitations.
With increasing $h_x$, FF states transit from gapped
one to gapless one. Such transition was first found in 2D in
Ref.~\refcite{FanWu2013PRL} and in 3D in Ref.~\refcite{LinDongArx}.
The gapless FF states are confirmed to be stable against phase fluctuations
in both 2D~\cite{YongBKT,Hu2014arXiv} and 3D~\cite{XJliu13PRA,Zheng2014TMatrix}.
On the other hand, the superfluids can transit from normal superfluids to
topological superfluids~\cite{Gong2011prl} in the presence of $h_z$ but without $h_x$.
With $h_x$, this transition is replaced by the transition from gapped FF
states to topological FF states. In the topological FF phase, there exist
band touching points (i.e. Weyl fermoins) as well as FF Cooper pairs
in the quasiparticle energy excitations.
Interestingly, a large region of the phase diagram is occupied by a gapless
topological FF superfluid with Weyl fermion excitations and $E_g<0$.
Instead, the gapless FF superfluids area becomes extremely small, which
separates the gapped FF states from the gapless topological FF superfluids.
In 2D, the parameter region where such gapless FF superfluids including both gapless FF and gapless
topological FF states exist becomes much smaller~\cite{Hu2014arXiv,YongBKT}, given that
the gap at $k=0$ begins increasing after its closing as $h_z$ becomes larger.

\subsubsection{Linear quasiparticle excitation spectrum}
Weyl fermions possess linear energy dispersions. To check this, we first consider
SO coupled Fermi gases subject to only out-of-plane Zeeman field $h_z$~\cite{Gong2011prl}.
The quasiparticle energy excitations can be analytically written as
\begin{equation}
E_{\pm}^{\lambda}({\bf k})=\lambda\sqrt{\xi_{\bf k}^2+\alpha^2k_\perp^2+h_z^2+|\Delta_0|^2 \pm
2\sqrt{h_z^2(\xi_{\bf k}^2+|\Delta_0|^2)+\alpha^2 k_\perp^2\xi_{\bf k}^2}}
\end{equation}
where $\lambda=\pm$ indicating the particle ($+$) and hole branches ($-$) respectively,
$\xi=\hbar^2{\bf k}^2/2m-\mu$, and $k_\perp=\sqrt{k_x^2+k_y^2}$.
Without SO coupling and Zeeman fields $(\alpha=h_z=0)$, it becomes the typical BCS
superfluids excitations $E_{{\bf k},\pm}^\lambda=\lambda\sqrt{\xi_{\bf k}^2+|\Delta_0|^2}$.
In the previous subsection, we have discussed that in 2D the topological phase transition happens at
the point across which the gap first closes and then reopens, generally signaling a
sharp change of Chern number. In 3D, we cannot define Chern number in the whole momentum
space. However, we can still find such gap close points. The kinetic energy
$\hbar^2 k_z^2/2m$ along the z direction can be incorporated into the chemical potential
so that the excitation gap becomes $|\sqrt{(\mu-\hbar^2 k_z^2/2m)^2+\Delta_0^2}-h_z|$. Clearly,
when $h_z=|\Delta_0|=h_c$ ($\mu>0$), the gap closes. However, the gap keeps closed even with
increasing $h_z$ since $k_z$ can vary. From $h_z=\sqrt{(\mu-\hbar^2 k_z^2/2m)^2+\Delta_0^2}$,
the gapless points exist at four points: $k_{z}=k_c=\pm\sqrt{\mu+\sqrt{h_z^2-|\Delta_0|^2}}$ and
$k_{z}=k_c=\pm\sqrt{\mu-\sqrt{h_z^2-|\Delta_0|^2}}$, when $|\Delta_0|<h_z<\sqrt{\mu^2+|\Delta_0|^2}$,
and only two points: $k_{z}=k_c=\pm\sqrt{\mu+\sqrt{h_z^2-|\Delta_0|^2}}$, when $h_z>\sqrt{\mu^2+|\Delta_0|^2}$.

Around the zero energy point ${\bf k}={\bf k}_W$ with ${\bf k}_W=(0,0,k_{c})$, the energy
dispersion can be approximately written as
\begin{eqnarray}
E_{-}^{\pm}(k_{x}=\delta k_{x},k_y=0,k_z=k_{c})&=&\pm\frac{\alpha\Delta_0}{h_z}|\delta k_x| ,
\label{ln_excitation1} \\
E_{-}^{\pm}(k_x=0,k_y=0,k_z=k_{c}+\delta k_z)&=&\pm\frac{\sqrt{2}\hbar^2|\xi_{{\bf k}_W}k_{c}|}
{m\sqrt{\Delta_0^2+\xi_{{\bf k}_W}^2}}\delta k_z \,
\label{ln_excitation2}
\end{eqnarray}
Clearly, this spectrum is linear along all three directions. The anisotropy of the
slope along the z direction and the plane originates from the plane SO coupling of atoms.
At the transition point where $k_c=0$ or $\xi_{{\bf K}_W}=0$, the liner behavior around the
z direction vanishes.

\begin{figure}[bt]
\centerline{\psfig{file=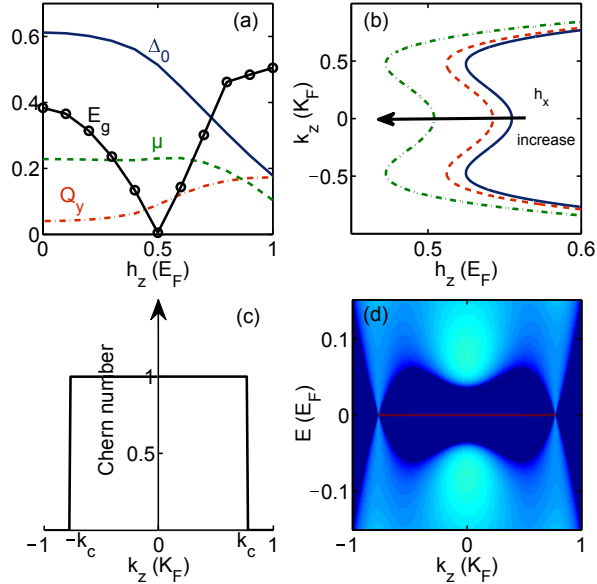,width=3.4in}}
\vspace*{8pt}
\caption{(Color online) (a) Plot of order parameter $\Delta _{0}$, $Q_{y}$,
chemical potential $\protect\mu$ and the quasiparticle excitation gap $%
E_{g}$ at $k_z=0$, as a function of $h_{z}$ with $h_{x}=0.2E_{F}$.
(b) Gap closing points ($E_{g}=0$) in the $(h_{z},k_{z})$
plane for $h_{x}=0$ (solid blue line), $h_{x}=0.1E_F$ (dashed red line), and $%
h_{x}=0.2E_F$ (dashed-dot green line). (c) Chern number for a fixed $k_{z}$
plane. (d) the density of states at $k_{y}=0$. The
light blue region and the red line (Fermi arch) represent bulk and surface
excitations, respectively. In (c) and (d), $h_{x}=0.2E_{F}$, $%
h_{z}=0.55E_{F} $. In all four figures $\protect\alpha K_{F}=E_{F}$, $%
1/a_{s}K_{F}=-0.1$. From Ref.~\protect\refcite{Yong2014PRL}.}
\label{TS_property}
\end{figure}

In the presence of in-plane Zeeman fields $h_x$, $Q_y\neq 0$, and
the gap at $k_{\perp}=0$ closes when
\begin{equation}
(h_{x}+\alpha Q_{y}/2)^{2}+h_{z}^{2}=(\hbar ^{2}k_{z}^{2}/2m-\bar\mu
)^{2}+\Delta _{0}^{2},
\label{zero_pts}
\end{equation}%
with $\bar\mu=\mu-Q_y^2/8m$. This equation determines the position $\mathbf{k}_{W}$ of
the Weyl nodes. When $h_x=Q_y=0$, it reduces to $h_z^2=(\hbar ^{2}k_{z}^{2}/2m-\mu
)^{2}+\Delta _{0}^{2}$, exactly the same as previous results with only $h_z$.
The gap at $k_z=0$ first closes and then reopens as shown in Fig.~\ref{TS_property}%
(a). Across the gap closing point, the order parameter $\Delta_0$ is still finite.
The finite $Q_y$ indicates the FF superfluids. Analogous to the system with pure $h_z$,
there are two regions: $h_{z}^2>
\bar\mu ^{2}+\Delta _{0}^{2}-(h_{x}+\alpha Q_{y}/2)^{2}$ with two zero excitations
and $\Delta _{0}^{2}-(h_{x}+\alpha Q_{y}/2)^{2}<h_{z}^2<\bar\mu%
^{2}+\Delta _{0}^{2}-(h_{x}+\alpha Q_{y}/2)^{2}$ and $\bar\mu>0$ with four zero excitations. Both
critical values for $h_{z}$ decrease with increasing $h_{x}$ as shown in
Fig.~\ref{TS_property} (b). As $h_x$ approaches such that $\Delta_0^2=(h_x+\alpha Q_y/2)^2$,
the topological transition occurs at $h_z=0$ as shown in Fig.~\ref{phase}. Fig.~\ref{TS_property}(b)
also implies that the properties of Weyl fermions, such as the number, position,
and creating and annihilating, can be readily tuned through changing Zeeman fields.

\begin{figure}[bt]
\centerline{\psfig{file=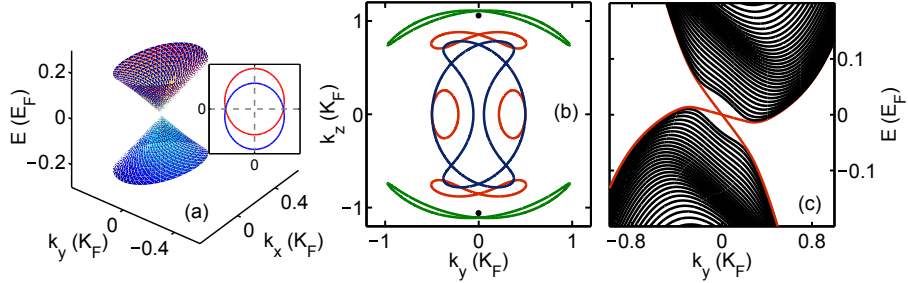,width=4.4in}}
\vspace*{8pt}
\caption{(Color online) (a) Quasiparticle excitations around the Weyl node $%
\mathbf{k}_{W}=\left( 0,0,0.77K_{F}\right) $ with $h_{z}=0.55E_{F}$ and $%
h_{x}=0.2E_{F}$. The contours of energy with $E=0.1E_{F}$ (
red line) and $E=-0.1E_{F}$ (blue line) are given in the inset.
(b) Contours of zero energy quasiparticle
spectrum in the plane $(k_{y},k_{z})$ with $k_{x}=0$. Here $h_{x}=0.5E_{F}$,
$h_{z}=0$ (blue line), $h_{z}=0.2E_{F}$ (red line), $h_{z}=E_{F}$ (green
line), and $h_{x}=0.2E_{F}$, $h_{z}=E_{F}$ (black points) correspond to
blue, red, green, black square points respectively in Fig.~\ref{phase}. There is no
zero energy excitations for the brown point in the gapped FF phase.
(c) Quasiparticle excitation spectrum in the gapless topological FF phase
($h_{x}=0.5E_{F}$ and $h_{z}=0.2E_{F}$) as a function of $k_{y}$ with
fixed $k_{z}=0.8K_{F}$ and confinement in the $x$ direction. The black lines
correspond to the bulk states while the red lines the surface states. Here
$\protect\alpha K_{F}=E_{F}$, $1/a_{s}K_{F}=-0.1$.}
\label{gasless_property}
\end{figure}

Fig.~\ref{gasless_property}(a) presents the linear excitations cove along the
$k_x$ and $k_y$ directions. The linear characteristic along the $k_z$ direction
can be seen from Fig.~\ref{TS_property}(d). It is important to note that the slope
along different directions is different, indicating that these Weyl fermions are
anisotropic. The difference along the z direction originates from the anisotropic SO
coupling, which only exists in the $(x,y)$ plane. Such difference also exists even
without $h_x$ as seen from Eq.~\ref{ln_excitation1} and Eq.~\ref{ln_excitation2}.
However, the difference between the slopes along x and y directions results
from finite momentum FF pairings.

Alternatively, the systems with Weyl fermions can be regarded as stacks of
quantum Hall insulators of quasiparticles in momentum space parameterized by $k_z$.
Weyl fermions emerge at the edges between quantum Hall insulators and normal
insulators, where Chern number changes sharply. In the topological FF phase,
because the quasiparticle excitations are gapped (except at the Weyl nodes)
in the 2D plane with a fixed $k_{z}$, we can calculate the Chern number for
the hole branch for each $k_{z}$ plane
\begin{equation}
C\left( k_{z}\right) =\frac{1}{2\pi }\sum_{n}\int dk_{x}dk_{y}\Omega
^{n}(k_{x},k_{y}),  \label{Chern_equ}
\end{equation}%
where $n$ is the index for hole branches, and the Berry curvature in the
z direction can be written as \cite{XiaoRMP}
\begin{equation}
\Omega ^{n}=i\sum_{n^{\prime }\neq n}\left[ \frac{\langle n|\partial
_{k_{x}}H_{B}|n^{\prime }\rangle \langle n^{\prime }|\partial
_{k_{y}}H_{B}|n\rangle -(k_{x}\leftrightarrow k_{y})}{(E_{n\mathbf{k}%
}-E_{n^{\prime }\mathbf{k}})^{2}}\right] ,
\end{equation}%
and $n^{\prime }$, which is not equal to $n$, runs over the eigenstates of $%
H_{B}$.

Fig.~\ref{TS_property}(c) shows that when $|k_z|<k_c$, Chern number $C=1$,
and $C=0$ otherwise. It is well know that there exist chiral edge states
between a quantum Hall insulator and a normal insulator. Thus, chiral edge
states are expected to exist if the edges are imposed along the x or y direction
for a fixed $k_z$ where Chern number is nonzero. For instance, with the
confinement along the x direction, the edge states spectrum should intersect
at $k_y=0$ and $E=0$. This results in a non-closed Fermi surface (i.e. Fermi arch)
connecting two Weyl points (shown in Fig.~\ref{TS_property}(d)),
which consists of surface states. The Fermi arch can also be regarded as
zero energy Majarona fermion flat band. We note that if the SO coupling
is the equal Rashba-Dresshaul type,
zero points form a loop in momentum space~\cite{Melo2012PRL}
and there exist a Majorana plane filling in the loop~\cite{Chunlei2014arXiv}.

\subsubsection{Gapless Topological Superfluids}
Topological FF superfluids are gapless and gapless excitations occur at
a pair of points ${\bf k}={\bf k}_W$ (i.e. Weyl points) where the
particle and hole branches touch. However, with increasing $h_x$,
the distortion of the quasiparticle excitations along the y direction
becomes sufficiently strong that the particle branch cuts the zero
energy plane, leading to gapless FF superfluids ($E_g<0$).
Note that the minimum of the particle branch and the maximum of the hole branch
do not generally occur at the same ${\bf k}$ because of the asymmetry induced
by $h_x$ corresponding to symmetry $\sigma_0\otimes i\sigma_z \mathcal{K}$ broken with
the complex conjugate operator $\mathcal{K}$.
However, the particle-hole symmetry is still conserved, meaning that when the minimum of
the particle branch occurs at ${-\bf k}$, the maximum of the hole branch occurs at
${\bf k}$. Such gapless FF superfluids can be topological or topological trivial,
depending on whether there exist a pair of zero energy points where the
particle and hole branches touch with the linear excitation dispersion
and whether such pair of zero energy points is connected.

Fig.~\ref{gasless_property}(b) displays the zero energy contour of each gapless states.
For topological FF states, there are two isolated points along the $k_z$ direction.
For gapless topological FF states, such gapless points become closed loops
connected at a Weyl point where the particle and hole branches touch. For gapless 
FF states (topological trivial),
we can divide them into two groups depending on whether the state possesses the
band touching points. Although the quasiparticle spectrum of one group has the
touching points with linear dispersions (see the blue line in Fig.~\ref{gasless_property}(b)),
it does not exhibit any topological properties. To confirm that the touching points 
in gapless topological FF superfluids are indeed Weyl fermions, we calculate the Chern
number of the hole branches and find that it is nonzero when $k_z$ lies
between those two points and zero otherwise. Fig.~\ref{gasless_property}(c)
illustrates the surface states (red line) with the confinement along the x direction.
Such surface states can still form the zero energy Fermi arch similar to
Fig.~\ref{TS_property}(d).

For an isotropic SO coupled Fermi gas subject to Zeeman fields, there
only exist gapless topological superfluids~\cite{Dong2014arXiv},
but not topological superfluids,
in the sense that the same Zeeman field not only generates Weyl points
but also creates FF pairings, in contrast to Rashba SO coupled Fermi
gas where the Zeeman field perpendicular to the SO coupling plane generates
Weyl points and the Zeeman field in the plane creates FF pirings.
The surface states are visualized in real space in Ref.~\refcite{Dong2014arXiv}.

\subsection{Sound Speeds as an Experimental Signal}
FF superfluids have finite center-of-mass momentum Cooper pairs. To
detect such exotic states in experiments, the direct method is to
probe the momenta of Cooper pairs. In cold atom systems, such finite
momenta can be reflected by the pair momentum distribution
after the time-of-flight expansion~\cite{Torma2007PRL,Meisner2007PRB}. Other indirect methods,
such as noise correlations~\cite{Torma2008PRA}, have also been suggested. However,
to date, in cold atom experiments, only the density profiles of both components
have been measured. This experiment only partially confirm the existence
of FFLO superfluids.

Sound modes which have linear energy dispersions are the collective
excitations excited by density fluctuations. The speeds of sound
are defined as the slopes of the linear dispersion, which reflect
the robustness of superfluidity in superfluids/superconductors.
For instance, in BECs, the critical velocity to destroy superfluidity
is determined by sound speeds. In the BCS superfluids, although
the speeds of sound are generally larger than the velocity
defined by the gap, they can still be experimentally
measured~\cite{Ketterle1997PRL,Joseph2007PRL} by
observing the propagation of a localized density perturbation
created by a laser beam. In this subsection, we suggest that the
speeds of sound can be employed as a signal~\cite{Torma2011PRA,He2012PRA}
reflecting the topological phase transition from
gapped FF superfluids to topological FF superfluids~\cite{Yong2014PRL}.

\begin{figure}[bt]
\centerline{\psfig{file=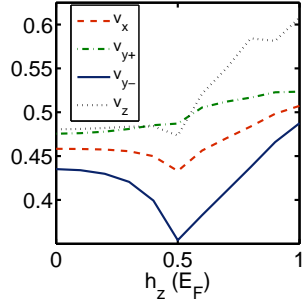,width=1.7in}}
\vspace*{8pt}
\caption{(Color online) Sound speeds as a function of $h_{z}$. Dashed red
and dotted black lines represent the speed along $x$ and $z$ directions.
Solid blue and dashed-dot green lines represent sound speeds along negative $%
y$ and positive $y$ directions, respectively. The unit of sound speed is $%
v_{F}=\hbar K_{F}/m$. Here $h_{x}=0.2E_{F}$, $\protect\alpha K_{F}=E_{F}$,
and $1/a_{s}K_{F}=-0.1$. From Ref.~\protect\refcite{Yong2014PRL}.}
\label{sound}
\end{figure}

In Fig.~\ref{sound}, we plot the speeds of sound along each direction with respect
to $h_z$ corresponding to the transition from gapped FF superfluids to
topological FF superfluids. The speeds of sound are anisotropic in all
different directions. The anisotropy between the z direction and the
$(x,y)$ plane comes from the SO coupling and the anisotropy in the
plane comes from the finite momentum pairings. They are anisotropic
even in the positive and negative y directions, reflecting the existence
of FF superfluids. In addition, there is a minimum in the speeds of sound at
the topological transition point where quantum fluctuations are relatively
strong. This signal may be used to measure the topological transition in
experiments.

\section{Conclusion}
In this review, we discussed the topological properties of FF superfluids in
both low dimensions and 3D. The mechanism of FF superfluids is
completely different from the traditional Zeeman field induced FFLO superfluids.
Here the FF superfluids originate from the symmetry breaking of the single
particle band structure induced by SO coupling and in-plane Zeeman fields.
In low dimensions, such FF superfluids can become topological when an
out-of-Zeeman field is included. In the topological FF superfluids,
Majorana fermions are locally accommodated in real space. The finite
BKT temperature paves the way for experimentally observing the Majorana
fermions as well as FF type pairings in 2D. In 3D, an out-of-Zeeman field
can still drive the system into a topological state. Compared with 2D, the topology
is essentially reflected by the emergence of Weyl fermions in momentum space.
Such Weyl fermions can be different from traditional ones in the sense that
the quasiparticle excitations become gapless except the Weyl points.
In experiments, the speeds of sound may provide an effective approach to
probe the topological FF states.

\section*{Acknowledgements}
We thank Z. Zheng, C. Qu and K. Sun for helpful discussion and critical reading.
Y. Xu and C. Zhang are supported by ARO(W911NF-12-1-0334), AFOSR (FA9550-13-1-0045),
and NSF-PHY (1249293).

\end{document}